\documentclass[english,a4paper]{article}
\usepackage[T1]{fontenc}
\usepackage[cp1250]{inputenc}
\usepackage{amssymb}
\usepackage{hyperref}
\usepackage{amsfonts}
\usepackage{graphicx}

\begin{document}

\title{A connection between optimal control theory and adiabatic passage techniques in quantum systems}
\author{E. Ass\'emat, D. Sugny\footnote{Laboratoire Interdisciplinaire Carnot de
Bourgogne (ICB), UMR 5209 CNRS-Universit\'e de Bourgogne, 9 Av. A.
Savary, BP 47 870, F-21078 DIJON Cedex, FRANCE,
dominique.sugny@u-bourgogne.fr}}

\maketitle

\begin{abstract}
This work explores the relationship between optimal control theory
and adiabatic passage techniques in quantum systems. The study is
based on a geometric analysis of the Hamiltonian dynamics
constructed from the Pontryagin Maximum Principle. In a
three-level quantum system, we show that the Stimulated Raman
Adiabatic Passage technique can be associated to a peculiar
Hamiltonian singularity. One deduces that the adiabatic pulse is
solution of the optimal control problem only for a specific cost
functional. This analysis is extended to the case of a four-level
quantum system.
\end{abstract}

\section{Introduction}
Adiabatic passage techniques are an efficient approach to design a
control pulse which drives a quantum system from an initial state
to a specific target state (for recent overviews, see
\cite{vitanov,guerin} and references therein). Such processes have
found a large variety of applications ranging from Nuclear
Magnetic Resonance \cite{MRI} and quantum information to atomic
and molecular excitations \cite{shapiro,rice}. Adiabatic methods
are usually achieved by using a series of intense pulses which can
be frequency chirped, the frequencies and the chirped being chosen
with respect to the structure of the energy levels of the quantum
system. The modification of the shape of the pulse envelope and
the chirping rate must be sufficiently slow so as to fulfill
adiabatic conditions. In this paper, we will consider one of the
most well-known adiabatic processes, the Stimulated Raman
Adiabatic Passage (STIRAP) excitation \cite{stirap,stirapN1},
which involves a counterintuitive sequence of two pulses in a
three-level quantum system. This approach, which allows a complete
population transfer between two energy levels, is also robust in
the sense that it is not sensitive to small variations of the
field parameters. The adiabatic methods have however some major
drawbacks, which can be roughly summarized in a high pulse energy
and a long control duration. Such constraints can be problematic,
e.g., if other concurrent physical or chemical processes with the
same time scale occur during the control. This statement can be
expressed by saying that adiabatic fields are not able to reach
the physical limits of the best possible performance of the system
in terms of duration or energy of the pulse. These limits can be
established by using optimal control theory
\cite{brif,rabitz,shapiro,rice}, which has been introduced in the
eighties in the quantum control community. An optimal control
problem can be solved either from geometric and analytic tools
\cite{boscain1,boscain2,sugny3,lapert,bernard1,bernard2,khaneja}
for systems of low dimensions or from numerical algorithms in the
case of large dimensional quantum systems
\cite{brif,rice,mono,grapegeneral}. The two approaches are based
on the application of the Pontryagin Maximum Principle (PMP)
\cite{pont,bonnard,jurdjevic}, which transforms the control
problem into an Hamiltonian one. In this framework, the
computation of the optimal solution is replaced by the
determination of an Hamiltonian trajectory satisfying given
boundary conditions. A general problem with optimal solutions is
their lack of robustness against variations of the system
parameters and of the control field. Note that a method based on
the simultaneous control of different systems can be used to
improve this latter property (see e.g. \cite{skinner}), but this
purely numerical approach does not give any insight into the
robustness nature of the optimal solutions.

In the present paper, we study the relationship between adiabatic
and optimal pulses in the case of a three-level quantum system
where the STIRAP solution can be used. This question has been
analyzed in different works over the past few years but not
directly from a geometric optimal control perspective. This work
complements therefore several earlier numerical
\cite{hornung,malinovsky,sola,zhang,boscain1,optadia1,optadia2,optadia3,sugny1,sugny2}
and analytical studies \cite{yuan,boscain1} on this relation. To
summarize some previous results, we mention that the authors of
\cite{optadia1} argue that the STIRAP technique cannot be
expressed as an optimal control solution. The counterintuitive
sequence of STIRAP was recovered numerically in \cite{sola} by
considering as cost functional the population of the intermediate
level. In Ref. \cite{boscain1}, a tracking technique aiming at
minimizing the population into the intermediate state was proposed
to recover the adiabatic solution, but no insight into the
relation between the adiabatic pulse and the optimal one was
given. In \cite{yuan}, the Hamilton-Jacobi Bellman theorem was
used to solve the optimal control problem.

In this work, we first consider the energy minimization control
problem. Using the Hamiltonian structure derived from the PMP, we
show that the STIRAP solution can be associated to a peculiar
singularity of this system. However, in this case, the adiabatic
pulse can be viewed as a singular limit solution of the
Hamiltonian system which cannot be approached smoothly by an
optimal control field. In a second step, we show that the STIRAP
pulse is solution of an optimal control problem corresponding to a
specific cost functional, that we call \emph{STIRAP cost}. Using
this geometric analysis, we extend this approach to the case of a
four-level quantum system \cite{stirapN1,stirapN2,stirapN3}.

The paper is organized as follows. We present the system in Sec.
\ref{sec2a} and we recall some basics about the STIRAP solution
and the application of the PMP in Sec. \ref{sec2b}. Section
\ref{sec2c} is devoted to the geometric analysis of the
Hamiltonian dynamics for the energy minimization problem. It is
shown in Sec. \ref{sec2d} that the adiabatic pulse is solution of
an optimal control problem for a specific cost functional. In Sec.
\ref{sec3}, we extend this study to the case of a four-level
quantum system. A summary is presented in Sec. \ref{sec4}. The
appendix \ref{app} collects some technical and mathematical
results.
\section{Optimal control of a three-level quantum
system}\label{sec2} \subsection{The model system}\label{sec2a} We
consider a three-level quantum $\Lambda-$system whose dynamics is
governed by the Schr\"odinger equation. The system is described by
a pure state $|\psi(t)\rangle$ belonging to a three-dimensional
Hilbert space $\mathcal{H}$ spanned by the basis
$\{|1\rangle,|2\rangle,|3\rangle\}$. The dynamics of the system is
controlled by the pump and the Stokes pulses which couple
respectively the states $|1\rangle$ and $|2\rangle$ and the states
$|2\rangle$ and $|3\rangle$. Note that there is no direct coupling
between the levels $|1\rangle$ and $|3\rangle$. The time evolution
of $|\psi(t)\rangle$ is given by
\begin{equation}\label{eq1}
i\frac{\partial}{\partial t} |\psi(t)\rangle =H(t)
|\psi(t)\rangle,
\end{equation}
where, in the interaction representation and the rotating-wave
approximation, the Hamiltonian $H(t)$ can be written as
\begin{equation}\label{eq2}
H(t)=\left(\begin{array}{ccc} 0 & u_1(t) & 0 \\
u_1(t) & -ik & -u_2(t) \\
0 & -u_2(t) & 0
\end{array}\right),
\end{equation}
with $u_1(t)$ and $u_2(t)$ representing the Rabi frequencies of
the pump and Stokes pulses, respectively. The two pulses are
assumed to be on-resonance with the corresponding frequency
transitions. Equation (\ref{eq1}) is written in units such that
$\hbar=1$. The parameter $k$ describes the relaxation rate of the
second level, which is the only state interacting with the
environment. Its role will be made clearer in the following. Note
that the Hamiltonian (\ref{eq2}) has already been considered in
other studies about the use of adiabatic passage techniques in
presence of dissipation \cite{yuan,vitanovnew,vitanov1,vitanov2}.
We denote by $c_1$, $c_2$ and $c_3$ the complex coefficients of
the state $|\psi(t)\rangle$, and we introduce the real
coefficients $x_i$ ($i\in \{1,2,\cdots,6\}$) defined by:
\begin{equation}\label{eq3}
c_1=x_1+ix_4,~c_2=x_5-ix_2,~c_3=x_3+ix_6.
\end{equation}
Using Eq. (\ref{eq1}), it is then straightforward to see that only
the coordinates $x_1$, $x_2$ and $x_3$ are coupled to each other.
This leads to the following differential system \cite{yuan}:
\begin{equation}\label{eq4}
\left(\begin{array}{lll} \dot{x}_1 \\ \dot{x}_2 \\ \dot{x}_3
\end{array}\right) =\left(\begin{array}{lll}
-u_1x_2 \\ u_1x_1-kx_2+u_2x_3 \\
-u_2x_2\end{array}\right),
\end{equation}
which reads in a more compact form as:
\begin{equation}\label{eq5}
\vec{x}=\vec{F}_0(\vec{x})+u_1\vec{F}_1(\vec{x})+u_2\vec{F}_2(\vec{x}),
\end{equation}
with $\vec{x}=(x_1,x_2,x_3)$, $\vec{F}_0(\vec{x})=(0,-kx_2,0)$,
$\vec{F}_1(\vec{x})=(-x_2,x_1,0)$ and
$\vec{F}_2(\vec{x})=(0,x_3,-x_2)$.

We aim at transferring the system from the state $|1\rangle$ to
the state $|3\rangle$ without losing population through the
dissipation of the state $|2\rangle$. In the new coordinates, this
corresponds to the passage for $x_1=1$ to $x_3=1$. This transfer
can be realized by a standard STIRAP strategy in which the pulses
are applied in a counterintuitive order, i.e. the Stokes pulse
precedes the pump pulse. We propose to revisit in the next section
this control problem by using the PMP.
\subsection{The Pontryagin Maximum Principle}\label{sec2b}
We first analyze the optimal control of this three-level system
with the constraint of minimizing the energy of the laser fields.
The total duration $T$ is fixed and there is no restriction on the
amplitudes of $u_1$ and $u_2$. For this energy minimization
problem, the cost functional $C$ can be written as:
\begin{equation}\label{eq6}
C=\int_0^T [u_1^2(t)+u_2^2(t)]dt,
\end{equation}
where $T$ is the fixed control duration. The PMP states that there
exist an adjoint state $\vec{p}=(p_1,p_2,p_3)$ and a negative
constant $p_0$ not simultaneously equal to zero such that the
optimal trajectory is solution of the following Hamiltonian
problem:
\begin{equation}\label{eq7}
\begin{array}{l}
H = \vec{p}\cdot (\vec{F}_0 + \sum_{i=1}^2 u_i\, \vec{F}_i) + p_0 (u_1^2(t)+u_2^2(t))\\
\dot{\vec{x}} = \frac{\partial H}{\partial \vec{p}}(\vec{x},\vec{p},v)\\
\dot{\vec{p}} = -\frac{\partial H}{\partial \vec{x}}(\vec{x},\vec{p},v)\\
H(\vec{x},\vec{p},v) = \textrm{max}_u H(\vec{x},\vec{p},u),
\end{array}
\end{equation}
where $'\cdot'$ stands for the scalar product between two vectors.
This optimal control problem with no dissipation, i.e. $k=0$, has
been already solved in Ref. \cite{boscain1} where it was shown
that the optimal solution is associated to an intuitive order of
the Stokes and pump pulses. Here, we consider the same analysis by
adding a dissipative term on the intermediate state which forces
the control to not populate this level, mimicking thus the
adiabatic trajectory. Since there is no constraint on the control
fields, these latter can be computed explicitly from the
maximization condition which leads to:
\begin{equation}\label{eq8}
\frac{\partial H}{\partial u_1}=0,~\frac{\partial H}{\partial
u_2}=0.
\end{equation}
In the regular case where $p_0$ can be normalized to $-1/2$, one
deduces that the optimal controls are given by $u_1=\vec{p}\cdot
\vec{F_1}$ and $u_2=\vec{p}\cdot \vec{F}_2$. Plugging the
expressions of $u_1$ and $u_2$ into the first equation of
(\ref{eq7}), it is straightforward to check that the Hamiltonian
$H$ reads:
\begin{equation}\label{eq9}
H=\vec{p}\cdot \vec{F}_0+\frac{1}{2}(\vec{p}\cdot
\vec{F}_1)^2+\frac{1}{2}(\vec{p}\cdot \vec{F}_2)^2.
\end{equation}
The corresponding Hamiltonian dynamics encodes all the information
about the optimal trajectories of the control problem. A global
overview of these solutions can be obtained through a geometric
analysis of this dynamics, which is made clearer by the
introduction of the following spherical coordinates:
\begin{equation}\label{eq10}
\left\{ \begin{array}{l}
    x_1 = r \sin \theta \cos \phi\\
    x_2 = r \cos \theta \\
    x_3 = r \sin \theta \sin \phi.
\end{array} \right.
\end{equation}
The dynamical system takes then the form:
\begin{equation}\label{eq11}
\left\{ \begin{array}{rcl}
    \dot{r} & = & - k r  \cos ^2 \theta\\
    \dot{\theta} & = & k \sin \theta \cos \theta - u_1 \cos \phi + u_2\sin \phi \\
    \dot{\phi} & = & \cot \theta (u_1 \sin \phi + u_2 \cos \phi).
\end{array} \right.
\end{equation}
It can be simplified by using the controls $v_1$ and $v_2$ such
that:
\begin{equation}\label{eq12}
\left\{ \begin{array}{rcl}
    v_1 & = & -u_1 \cos \phi + u_2 \sin \phi\\
    v_2 & = & -u_1 \sin \phi - u_2 \cos \phi.
\end{array} \right.
\end{equation}
Note that the cost $C$ is not modified since
$v_1^2+v_2^2=u_1^2+u_2^2$. The Hamiltonian $H$ reads now:
\begin{equation}\label{eq13}
H=-kr\cos^2\theta p_r+p_\theta
(k\sin\theta\cos\theta+v_1)-v_2p_\phi\cot
\theta-\frac{1}{2}(v_1^2+v_2^2).
\end{equation}
Using the maximization condition, one arrives at:
\begin{equation}\label{eq14}
v_1=p_\theta,~v_2=-\cot\theta p_\phi.
\end{equation}
Plugging the expressions of $v_1$ and $v_2$ into $H$ leads to:
\begin{equation}\label{eq15}
H=-kr\cos^2\theta p_r+k\cos\theta\sin\theta p_\theta
+\frac{1}{2}p_\theta^2+\frac{1}{2}\cot^2\theta p_\phi^2.
\end{equation}
The Hamiltonian dynamics is then characterized by the following
differential equations:
\begin{equation}\label{eq16}
\left\{ \begin{array}{rcl}
    \dot{r} & = & - k r  \cos ^2 \theta\\
    \dot{\theta} & = & k \sin \theta \cos \theta + p_\theta \\
    \dot{\phi} & = & \cot^2 \theta p_\phi \\
    \dot{p}_r & = & kp_r\cos^2\theta \\
    \dot{p}_\theta & = & -k r
    p_r\sin(2\theta)-k\cos(2\theta)p_\theta+p_\phi^2\frac{\cos\theta}{\sin^3\theta}
    \\
    \dot{p}_\phi & = & 0.
\end{array} \right.
\end{equation}
In the STIRAP process, the coordinate $x_2(t)$ remains equal to 0
for any time $t$, which avoids the dissipation effects and
maximizes the transfer to the state $|3\rangle $. In the spherical
coordinates, we have $\theta(t)=\pi/2$ and $\dot{\theta}(t)=0$
which leads to $p_\theta(t)=\phi(t)=0$, i.e. no motion is
possible. This remark is an illustration of the well-known fact
that the exactly adiabatic trajectory is singular and needs an
infinite transfer time.
\subsection{Geometric analysis of the Hamiltonian dynamics}\label{sec2c}
This preliminary analysis does not give any insight into the
geometric character of the adiabatic solution, and a more
intricate understanding of the Hamiltonian dynamics is required.
This can be done first by proving that the Hamiltonian given by
Eq. (\ref{eq15}) is Liouville integrable \cite{Arnold}, i.e. it
has as many constants of the motion as degrees of freedom. Since
$H$ does not depend on $\phi$, we can check that $p_\phi$ is a
constant of motion. Another constant of the motion can be found by
noting from Eq. (\ref{eq16}) that the product $rp_r$ is constant.
Using the change of coordinates $\rho=\log r$ and introducing the
new set $(\rho,p_\rho)$ of canonical coordinates \cite{Arnold},
one deduces that the Hamiltonian $H$
 becomes:
\begin{equation}\label{eq17}
H=-kp_\rho\cos^2\theta+k\sin\theta\cos\theta p_\theta
+\frac{1}{2}(p_\theta^2+\cot^2\theta p_\phi^2),
\end{equation}
and does not depend on $\rho$. The functions $H$, $p_{\phi}$ and
$p_{\rho}$ are three constants of motion and the system is
therefore Liouville-integrable.

\begin{figure}[htbp]
   \centering
     \includegraphics[scale=0.15]{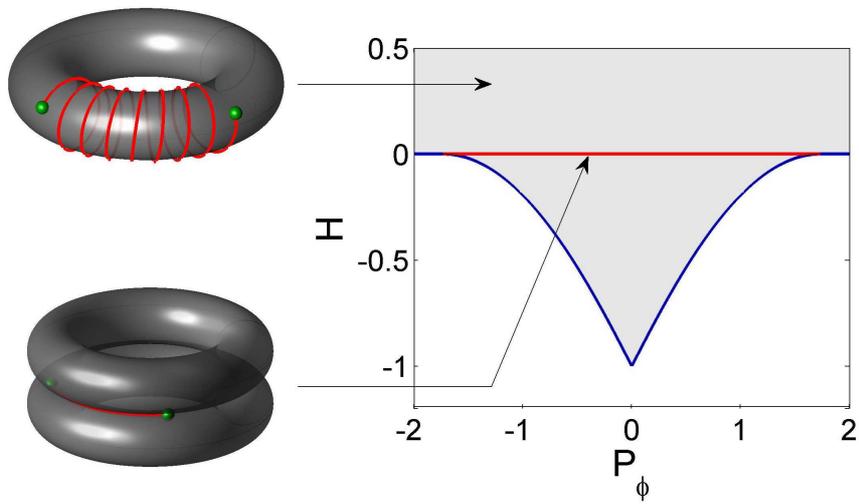}
  \caption{(Color Online) Image of the energy-momentum map
  $\mathcal{F}$ (gray) for the Hamiltonian of Eq. (\ref{eq15}).
The singular points are represented by solid blue (black) and red
(dark gray) lines. The red line indicates the positions of the
image of the bitori (see the text for details). Two trajectories
are schematically represented on a regular and on a singular tori
(the dots represent the initial and final points of the
trajectory). The upper trajectory represents a standard
oscillating solution obtained with the energy-minimum cost. The
lower trajectory is the ideal adiabatic trajectory, that can not
be reached with the energy-minimum cost. Numerical values are
taken to be $p_\rho=k=1$. The different quantities are unitless in all the figures.}
  \label{fig2}
\end{figure}

To simplify the description of the Hamiltonian dynamics, we will
consider in the following only a two-dimensional degree of freedom
system in the coordinates $(\theta,\phi)$ taking $p_\rho$ as a
parameter. This is possible since the dynamics does not depend on
$\rho$. The Liouville-Arnold theorem tells us that a given
trajectory defined by some initial conditions will curl around a
torus in the phase space \cite{Arnold,cushman}. This torus can be
either regular or singular, the former structure giving periodic
solutions while the latter leads to trajectories with an infinite
period. For one-dimensional Hamiltonian systems, note that the
singular tori are the well-known separatrices. Hence, one could
expect that the STIRAP solution is linked to such singular tori,
which is therefore crucial to characterize. For that purpose, we
consider the so-called \textit{energy-momentum map}:
$$
\mathcal{F}_{p_\rho} : (\theta,\phi,p_{\theta},p_{\phi}) \mapsto
(H,p_{\phi}),
$$
i.e. the set of all the possible values of the constants of the
motion $H$ and $p_\phi$, which is parameterized by the constant
$p_\rho$. A rigorous application of the Liouville-Arnold theorem
gives that the inverse image $\mathcal{F}^{-1}(H,p_\phi)$ of the
point $(H,p_\phi)$ of the energy-momentum set is a two-dimensional
regular torus if the two gradient vectors of the phase space
$\vec{\nabla} H=(\frac{\partial H}{\partial
p_\theta},\frac{\partial H}{\partial p_\phi},\frac{\partial
H}{\partial \theta},\frac{\partial H}{\partial \phi})$ and
$\vec{\nabla}p_\phi=(0,1,0,0)$ are not parallel for any point of
the torus. If the two vectors are parallel for some points, then
the pre-image is no longer a regular torus, but a singular one.
The topology of the singular torus can be of different types: a
point (for an equilibrium), a circle (for a periodic orbit) or a
more complicated two-dimensional structure. This latter geometry
can be determined from the singular reduction theory, which is
applied for this example in the appendix \ref{app}. Here, we only
compute the position of the different singular points in the
energy-momentum diagram. If the two gradients $\vec{\nabla} H$ and
$\vec{\nabla} p_\phi$ are parallels then the matrix defined by
\begin{equation}\label{eq18}
\left(\begin{array}{cc}
k \sin \theta \cos \theta + p_{\theta} & 0\\
\cot^2\theta\,p_{\phi} & 1\\
    k p_{\rho} \sin(2\theta) + k \cos(2\theta) p_{\theta} - \frac{\cos \theta}{\sin^3\theta} p_{\phi}^2 & 0\\
    0 & 0\\
\end{array}\right)
\end{equation}
is not of full rank, i.e. equal to 2 \cite{cushman}. We obtain two
different cases. In the first one, which corresponds to the STIRAP
process, we have $(\theta = \pi/2, p_{\theta} = 0)$ and $H=0$. In
the second situation, one arrives at:
$$
\begin{array}{l}
p_{\theta} = -k \sin \theta \cos \theta \\
p_{\phi} =\pm \sqrt{\frac{\sin^3\theta}{\cos \theta} (k p_{\rho}
\sin(2\theta) + k \cos(2\theta) p_{\theta})},
\end{array}
$$
which gives the equation of the boundary of the energy-momentum
set. The corresponding energy-momentum diagram is represented in
Fig. \ref{fig2}. Its qualitative structure does not depend on the
value of $p_\rho$.

\begin{figure}[htbp]
\begin{center}
\includegraphics[scale=0.7]{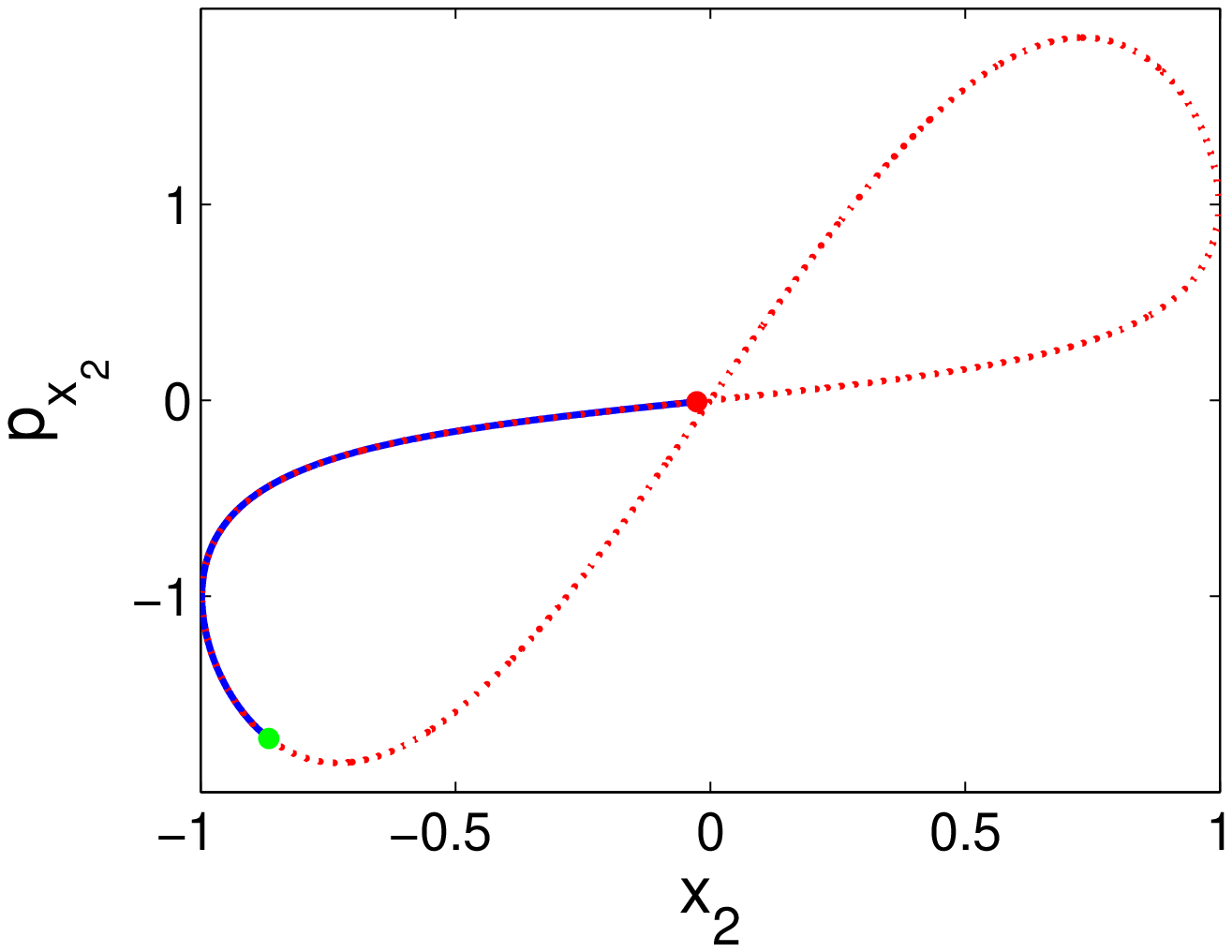}
\includegraphics[scale=0.3]{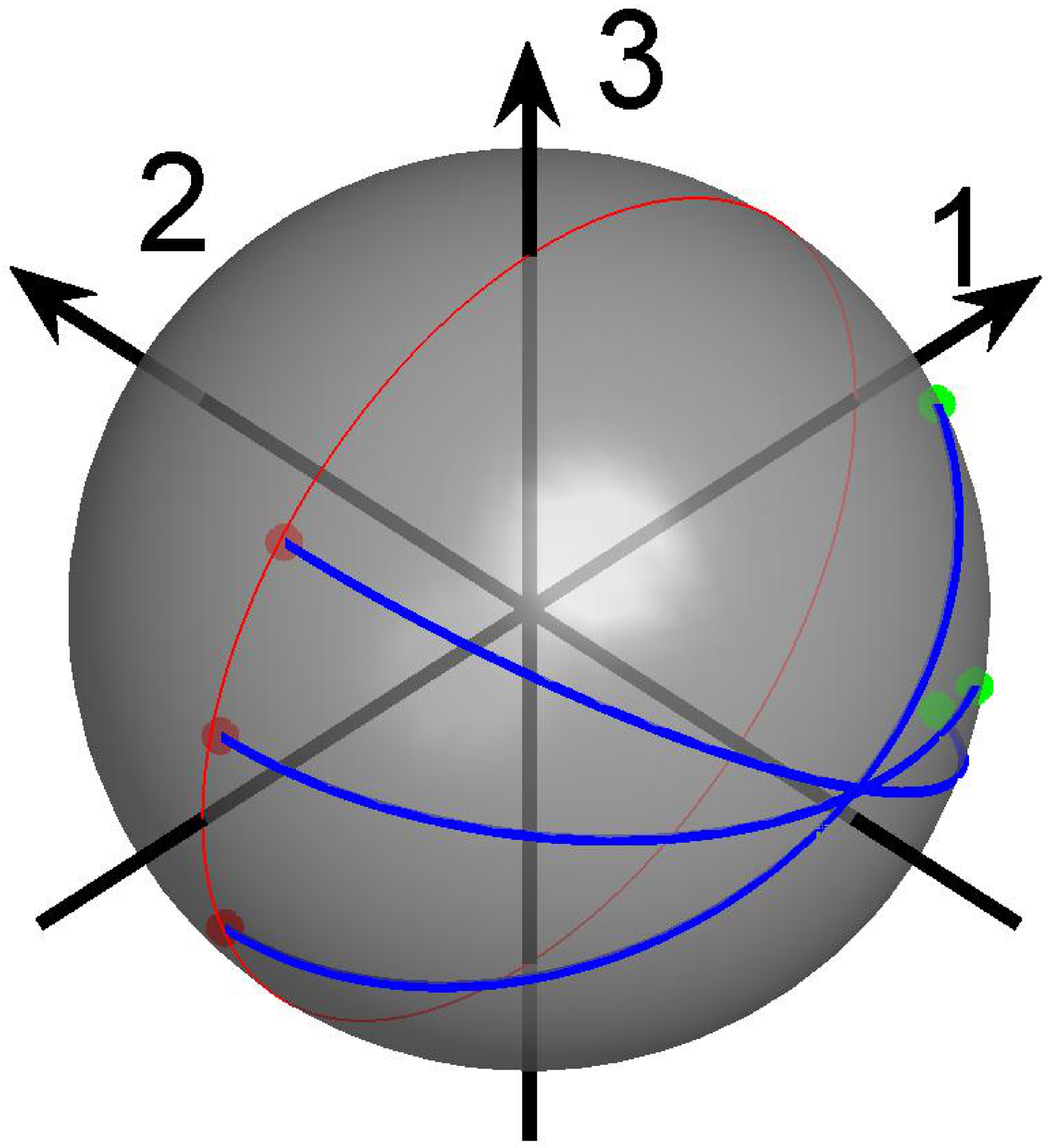}
\end{center}
\caption{(Color Online) Top: Projection in the $(x_2,p_{x_2})$
plane of a trajectory on the bitorus for $H = 0$, $p_{\phi} =
0.1$, $p_\rho = 1$, in the case where $\theta(0) \neq 0$. The
parameter $k$ is fixed to 1. Bottom: Projections onto the sphere
of different trajectories lying on a bitorus. The green (light
gray) and red dots (dark gray) are respectively the initial and
final positions of the trajectories. The great circle represents
the quantum states with no population in the intermediate level
$|2\rangle$.}\label{fig3}
\end{figure}

As mentioned above, a singular reduction \cite{cushman} allows to
make more precise the nature of this singularity (see the appendix
\ref{app} for details). In the energy-momentum diagram depicted in
Fig. \ref{fig2}, each point of the horizontal singular line of
equation $H=0$ is associated to a bitorus in the phase space, i.e.
two tori glued together along a singular circle. Straightforward
computations show that the points of this circle satisfy $x_2 =
p_{x_2} = 0$, i.e. the equation of the STIRAP solution. For
comparison, two schematic trajectories on a regular torus and on a
bitorus have been plotted in Fig. \ref{fig2}. We notice the
difference of structure between these two solutions. The set of
all the trajectories belonging to a given bitorus can also be
characterized. Two different cases can be distinguished. In the
first one, $\theta = \frac{\pi}{2}$, then $p_{\theta} = 0$ and the
system is on the singular circle which means that no motion is
possible. In the second subset, $\theta(t)\neq \frac{\pi}{2} $ at
some time $t$, but we observe numerically that the trajectory
returns at longer times to the singular circle such that $\theta =
\frac{\pi}{2}$. This point is illustrated in Fig. \ref{fig3}.
\begin{figure}[htbp]
\centering
\includegraphics[scale=0.4]{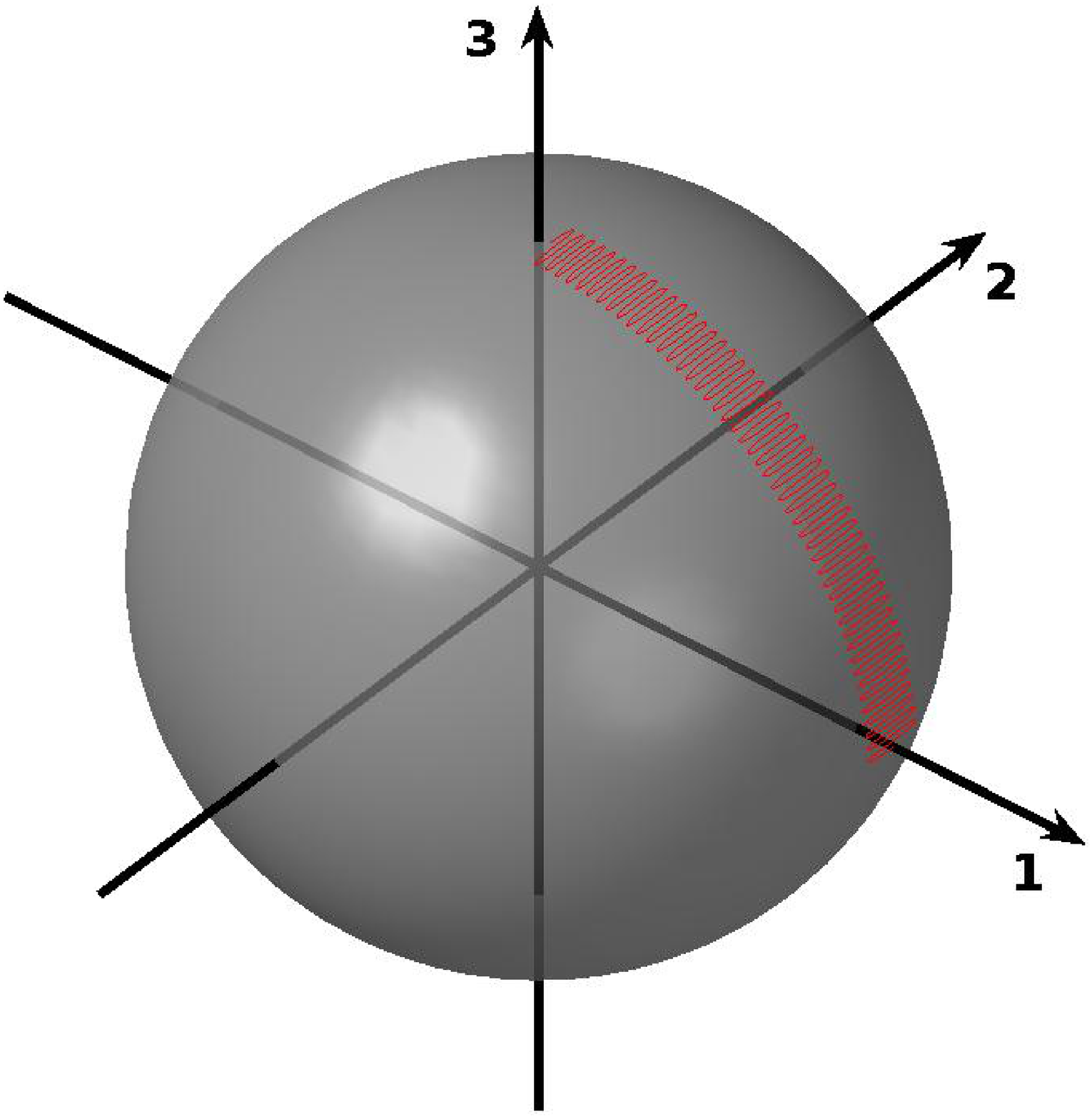}
\includegraphics[scale=0.7]{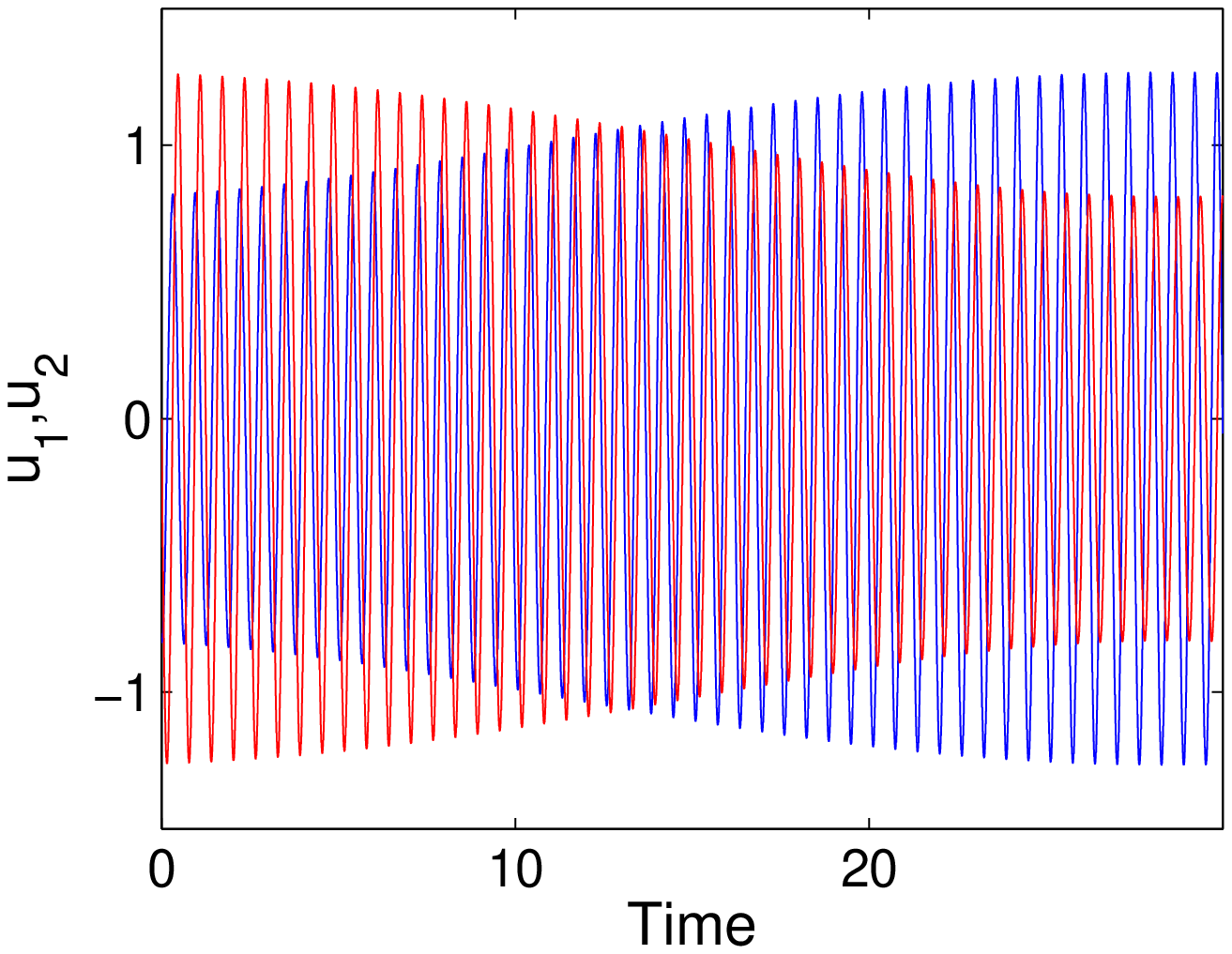}
\caption{(Color Online) Projection onto the sphere of an optimal
trajectory for the energy minimization control problem (top panel)
and of the corresponding fields (bottom panel). The pump and the
Stokes fields are respectively depicted in blue (black) and in red
(dark gray). The parameters are $ k = 1$ and $T=30$. The values of
the adjoint states are $p_\phi=15$, $p_\rho=69$ and the
Hamiltonian $H$ of Eq. (\ref{eq13}) is equal to 0.33.} \label{fig1}
\end{figure}

The last step of this preliminary study consists in solving
numerically the optimal control problem. For that purpose, we
choose some initial values of the momenta $p_\rho$, $p_\theta$ and
$p_\phi$ and we propagate the Hamiltonian equations (\ref{eq16})
during the time $T$. The distance to the target state is computed
at this final time. An example of solution is represented in Fig.
\ref{fig1}. Note the oscillatory behavior of the extremal
solution, which is far from the expected smooth and monotonic
evolution of the adiabatic pulse. These oscillations do not allow
to limit the losses from the intermediate state and the final
population of the state $|3\rangle$ is only of $0.82$. Such a
solution belongs to a regular torus close to the singular one. In
spite of this proximity in the energy-momentum diagram, two
completely different behaviors are obtained. In addition, a
systematic numerical analysis shows that the frequency of the
oscillations increases if the control amplitude increases as can
be seen in Tab. \ref{tab1}. We observe that the population
transfer is better, but the control frequency is larger. On the
other hand, if one tries to reduce the oscillatory character of
the solution by staying on the singular line or close to it then
it is not possible to reach the target state with efficiency, as
illustrated by the first line of Tab. \ref{tab1}. We thus conclude
that the limit of high energy does not allow to recover the STIRAP
process. The other intuitive limit would be a longer control
duration, but this is not favorable to STIRAP, as can be seen by
comparing the second and third lines of Tab. \ref{eqTab}. The
STIRAP process is therefore not an intrinsic optimal solution of
this problem. A peculiar cost functional has to be chosen in order
to highlight the optimal properties of the STIRAP technique.
\begin{table}\caption{\label{tab1}  Characteristics of different extremal solutions. The \emph{Freq.} and \emph{Amp.} columns indicate respectively a frequency and an amplitude averages of the fields $u_1$ and $u_2$.}
\begin{center}
\begin{tabular}{|c|c|c|c|c|c|c|}
\hline
$H$ & $p_{\phi}$ & $T$ & \textrm{Freq.} & \textrm{Amp.} & $C$ & $x_3^2(T)$\\
\hline
\hline
0 & 0.1 & 4 & 0 & 1.7 & 2.7 & $10^{-7}$\\
\hline
0.33 & 15 & 30 & 1.5 & 1.26 & 33.9 & 0.82\\
\hline
0.4 & 45 & 8 & 1.5 & 4.23 & 73.6 & 0.93\\
\hline
4.6 & 30 & 10 & 50 & 9.81 & 94.3 & 0.98\\
\hline
\end{tabular}
\label{eqTab}
\end{center}
\end{table}

\subsection{The STIRAP cost}\label{sec2d}
The study presented above makes clear that the energy minimum cost
is not well suited to the STIRAP process. The objective of this
section will be to determine a cost, which could force the optimal
solution to follow the STIRAP trajectory. From a dynamical point
of view, a basic argument is to prevent the Hamiltonian trajectory
to curl around a torus. This goal can be achieved by reducing the
oscillations around $x_2=0$, which leads in spherical coordinates
to the cost:
\begin{equation}\label{eq19}
C = \int_0^T\dot{\theta}^2dt = \int_0^T(k \sin \theta \cos \theta
+ v_1)^2dt.
\end{equation}
This choice also avoids populating the state $|2\rangle$ as shown
below. The Hamiltonian of the PMP has the form:
\begin{eqnarray*}
& & H = -k r \cos ^2 \theta p_r + k \cos \theta \sin \theta p_{\theta} + v_1 p_{\theta}\\
& & - v_2 \cot \theta p_{\phi} - \frac{1}{2}(k \sin \theta \cos
\theta + v_1)^2.
\end{eqnarray*}
The cost $C$ depending on $v_1$ and not on $v_2$, we choose to
optimize the Hamiltonian $H$ only with respect to $v_1$.
Optimizing also $v_2$ would lead once again to a static solution.
We obtain $v_1 = p_{\theta} - k \sin \theta \cos \theta$ and
plugging this expression into the Hamiltonian $H$ gives:
\begin{equation}\label{eq20}
H = -k r p_r \cos^2 \theta + \frac{1}{2} p_{\theta}^2 - v_2 \cot
\theta p_{\phi},
\end{equation}
where $v_2$ is a given time-dependent function. The equations of
motion become:
\begin{equation}\label{eq20a}
\left\{ \begin{array}{rcl}
    \dot{r} & = & - k r  \cos ^2 \theta\\
    \dot{\theta} & = & p_{\theta} \\
    \dot{\phi} & = & -\cot \theta \, v_2 \\
    \dot{p_r} & = &  k p_r  \cos ^2 \theta\\
    \dot{p_{\theta}} & = & - 2 k r p_r \sin \theta \cos \theta - \frac{p_{\phi}}{\sin ^2 \theta} v_2 \\
    \dot{p_{\phi}} & = & 0.
\end{array} \right.
\end{equation}
From Eq. (\ref{eq20a}), it is clear that the solution
$p_{\theta}(t) = 0$ for any time $t$ minimizes the cost $C$. As a
consequence, we get that $\theta(t)$ is constant along this
trajectory and the expression of the second control field $v_2$
comes from $\dot{p}_\theta = 0$:
\begin{equation}\label{eq21}
v_2 = - \frac{2 k r p_r \sin^3 \theta \cos \theta}{p_{\phi}}.
\end{equation}
Here the functions $r p_r$, $p_{\phi}$ and $\theta$ are constants
of motion, $v_2$ is also constant. Using Eq. (\ref{eq21}), one
gets the extremum value of $H$, which reads as:
\begin{equation}\label{eq21a}
H=2krp_r\cos^2\theta (\sin^2\theta-\frac{1}{2}).
\end{equation}
The STIRAP solution can be recovered from the limit case
$\theta\to \pi/2$. In this limit, we get $H=0$ and $v_1\simeq 0$,
leading to $u_1\cos\phi=u_2\sin\phi$ and
\begin{equation}\label{eq21b}
\frac{u_2}{u_1}=\frac{x_1}{x_3},
\end{equation}
which is characteristic of a STIRAP sequence.

To numerically solve this control problem, one should also note
that there exist some constraints on the initial momenta. Indeed,
if we denote by $\tau$ the typical relaxation time, the following
condition holds on the final time: $T << \tau$. As $\tau =
\frac{1}{k \cos^2 \theta}$ and $T = \left| \frac{\pi}{2 v_2 \cot
\theta}\right|$, this leads to the relation:
$$
|\frac{\pi p_{\phi}}{4 r p_r \sin^2 \theta}| \ll 1.
$$

\begin{figure}[htbp]
\centering
\includegraphics[scale=0.2]{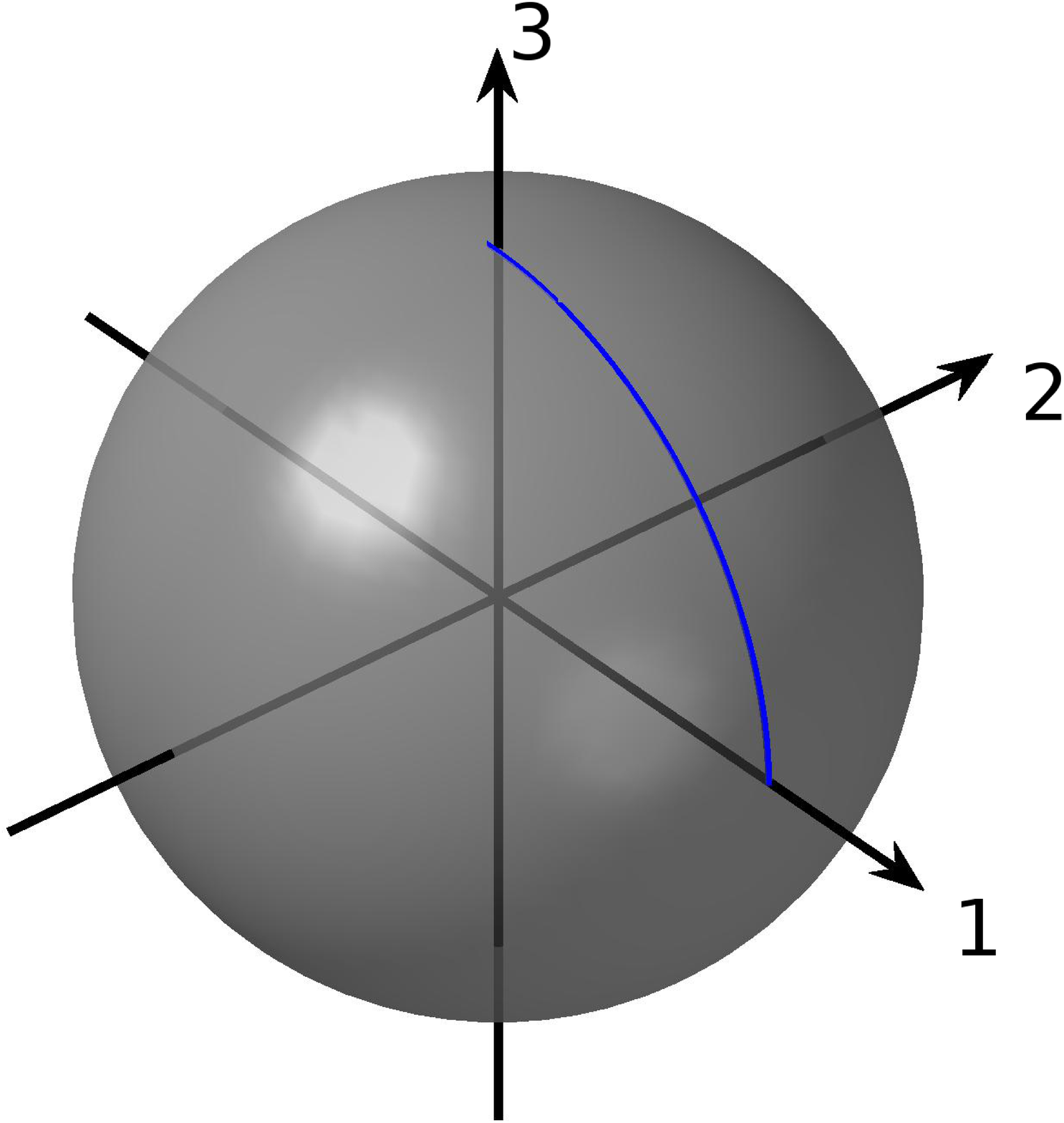}
\includegraphics[scale=0.7]{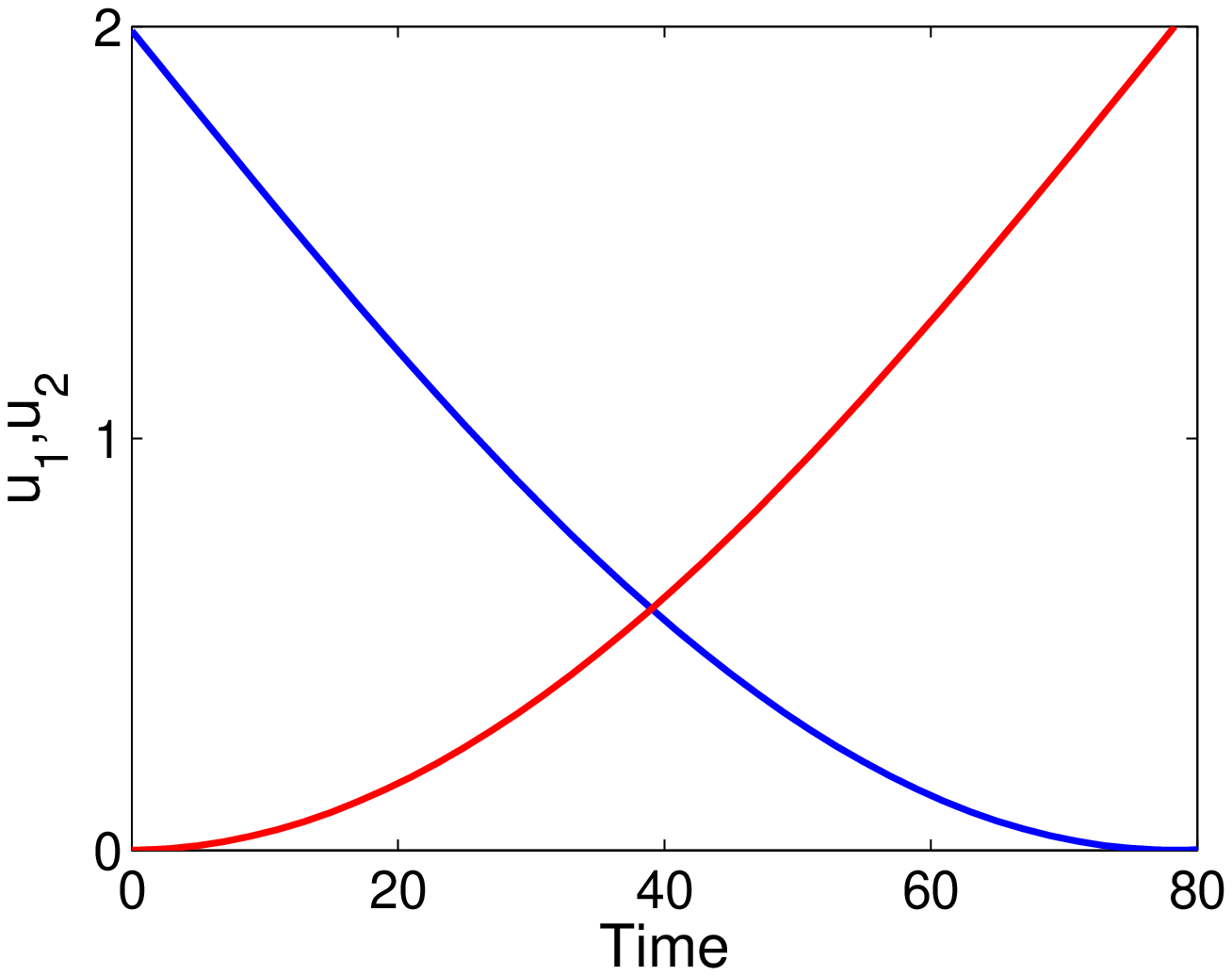}
\caption{(Color online) Optimal control fields obtained with the
Stirap cost (bottom) and the projection of the corresponding
trajectory onto the sphere $x_1^2+x_2^2+x_3^2=1$ (top). In the
bottom panel, the Stokes and the Pump pulses are respectively
depicted in blue (black) and red (gray). Numerical parameters are
taken to be $k=1$ and $T=80$. The initial momenta are
$p_\theta=0$, $p_\phi=0.1$, $p_\rho=10$. The Hamiltonian $H$ of
Eq. (\ref{eq20}) is equal to $10^{-3}$.} \label{fig4}
\end{figure}
An example of optimal trajectory is given in Fig. \ref{fig4}. Note
the counterintuitive order of the two control fields, one of the
main features of a STIRAP process. We can therefore conclude that
a STIRAP-like solution can be designed from the PMP if the right
cost is considered.
\section{Extension to a four-level quantum system}\label{sec3}
\begin{figure}[htbp]
\centering
\includegraphics[scale=0.7]{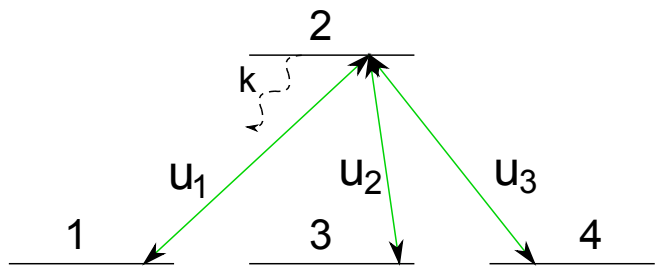}
\caption{(Color Online) The four-level tripod quantum system.
$u_1$, $u_2$ and $u_3$ stand for the pump and the two Stokes
pulses respectively. $k$ is the dissipation rate of the second
level $|2\rangle$.} \label{fig5new}
\end{figure}

We apply the same reasoning to a four-level quantum system with a
tripod structure
\cite{vitanov2,tripod1,tripod2,tripod3,tripod4,stirapN1,stirapN2,stirapN3}
as illustrated in Fig. \ref{fig5new}.

The dynamical evolution of this system is described by the
Hamiltonian:
\begin{equation}\label{eq22}
H = \left( \begin{array}{cccc}
    0 & - u_1 & 0 & 0 \\
    u_1 & -ik & - u_2 & - u_3\\
    0 & u_2 & 0 & 0 \\
    0 & u_3 & 0 & 0 \\
\end{array}\right)
\end{equation}
where $u_1$, $u_2$ and $u_3$ stand for the pump and the two Stokes
pulses respectively. The aim of the control is to transfer the
population of state $|1\rangle$ to a superposition of states
$|3\rangle$ and $|4\rangle$, while preventing the system to
populate the state $|2\rangle$. We first consider the energy
minimization cost $C$ given by
\begin{equation}\label{eq22a}
C=\frac{1}{2}\int_0^T[u_1^2(t)+u_2^2(t)+u_3^2(t)]dt.
\end{equation}
The pseudo-Hamiltonian of the PMP reads:
\begin{equation}\label{eq23}
\begin{array}{rcl}
    H & = &  -k x_2 p_{x_2} + u_1(x_1 p_{x_2} - x_2 p_{x_1}) + u_2 (x_2 p_{x_3} - x_3 p_{x_2})\\
     & & + u_3 (x_2 p_{x_4} - x_4 p_{x_2}) -\frac{1}{2}[u_1^2(t)+u_2^2(t)+u_3^2(t)].
\end{array}
\end{equation}
Introducing the spherical coordinates:
$$
\left\{ \begin{array}{l}
    x_1 = r \cos \theta_1 \sin \theta_2\\
    x_2 = r \cos \theta_2 \\
    x_3 = r \sin \theta_1 \sin \theta_2 \cos \theta_3\\
    x_4 = r \sin \theta_1 \sin \theta_2 \sin \theta_3
\end{array} \right. ,
$$
\begin{figure}[htbp]
\centering
\includegraphics[scale=0.7]{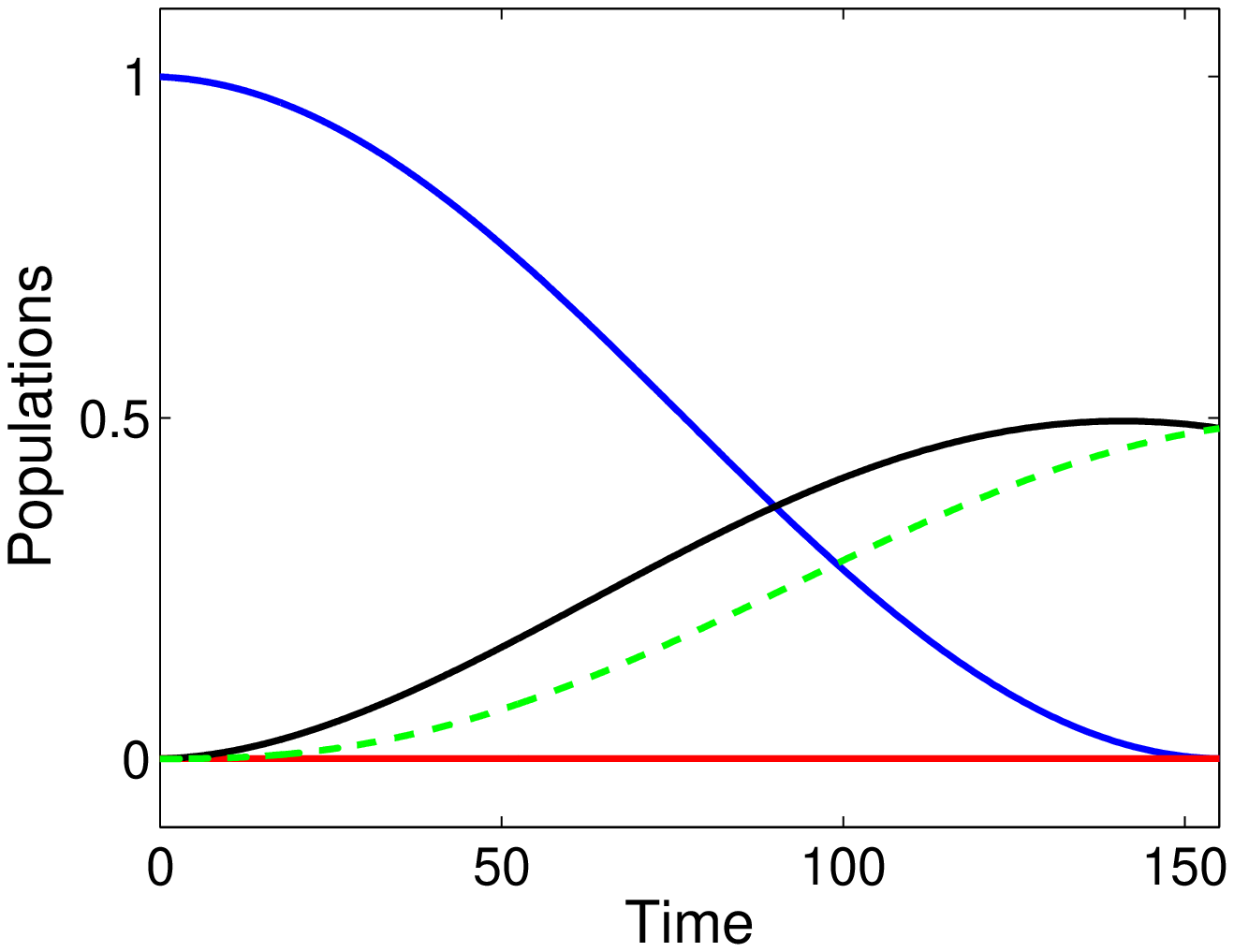}
\includegraphics[scale=0.7]{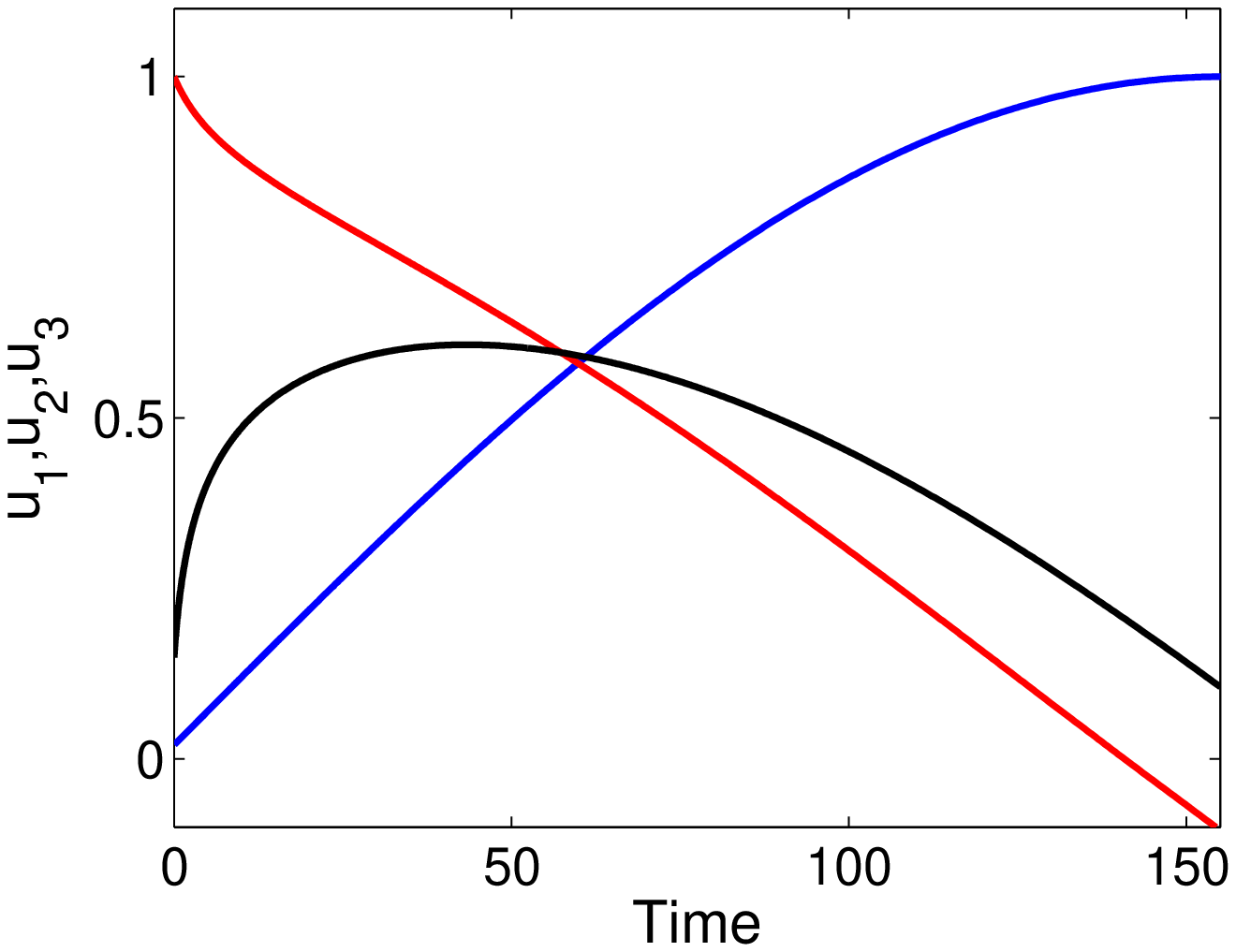}
\caption{(Color Online) Example of an optimal solution mimicking
the adiabatic evolution and leading to a superposition of the
states $|3\rangle$ and $|4\rangle$. The top panel depicts the
evolution of the population of the different states $|1\rangle$,
$|2\rangle$, $|3\rangle$ and $|4\rangle$ respectively in blue
(dark gray), red (light gray), black and green (dashed line). In
the bottom panel, the Pump, Stokes 1 and Stokes 2 pulses are
represented in blue (dark gray), red (light gray) and black,
respectively. Numerical values are $p_{\rho} = 100, w_1 = k = 1$.}
\label{fig5}
\end{figure}
the dynamical system takes the form:
\begin{equation}\label{eq24}
\left\{ \begin{array}{rcl}
    \dot{r} & = & - k r  \cos ^2 \theta_2\\
    \dot{\theta}_1 & = & u_1 \sin \theta_1 \cot \theta_2 + u_2 \frac{\cos \theta_2 \cos
    \theta_1 \cos \theta_3}{\sin \theta_2}\\
    & & + u_3 \cot \theta_2 \cos \theta_1 \sin \theta_3\\
    \dot{\theta}_2 & = & k \sin \theta_2 \cos \theta_2 - u_1 \cos \theta_1 + u_2 \sin \theta_1 \cos \theta_3\\
    & & + u_3 \sin \theta_1 \sin \theta_3 \\
    \dot{\theta}_3 & = & -u_2 \frac{\cos \theta_2 \sin \theta_3}{\sin \theta_1 \sin \theta_2}
    + u_3 \frac{\cos \theta_2 \cos \theta_3}{\sin \theta_1 \sin \theta_2}
\end{array} \right. .
\end{equation}
Using the following rotations on the control fields:
\begin{equation}\label{eq25}
\left\{ \begin{array}{rcl}
    v_2 & = & u_2 \cos \theta_3 + u_3 \sin \theta_3\\
    v_3 & = & -u_2 \sin \theta_3 + u_3 \cos \theta_3 \\
\end{array} \right. ,
\end{equation}
and
\begin{equation}\label{eq26}
\left\{ \begin{array}{rcl}
    w_1 & = & u_1 \sin \theta_1 + v_2 \cos \theta_1\\
    w_2 & = & -u_1 \cos \theta_1 + v_2 \sin \theta_1 \\
\end{array} \right. ,
\end{equation}
the Hamiltonian $H$ becomes:
\begin{equation}\label{eq27}
\begin{array}{rcl}
    H & = &  -k \, r \cos ^2 \theta_2 p_r + k \cos \theta_2 \sin \theta_2 p_{\theta_2} + w_1 \cot \theta_2 p_{\theta_1}\\
    & & + w_2 p_{\theta_2} + v_3 \frac{\cot \theta_2}{\sin \theta_1} p_{\theta_3}
    -\frac{1}{2}[w_1^2(t)+w_2^2(t)+v_3^2(t)].
\end{array}
\end{equation}
The analysis of the different Hamiltonian trajectories used in
Sec. \ref{sec2} can be done along the same lines. It can be shown
that the Hamiltonian $H$ is integrable since it possesses four
constants of the motion: $H$, $L_1 = x_3 P_{x_4} - x_4 P_{x_3}$,
$L_3 = x_1 P_{x_4} - x_4 P_{x_1} $ and $L_4 = x_1 P_{x_3} - x_3
P_{x_1}$. Here again, we obtain that the adiabatic trajectory lies
on a singular torus in the phase space. In this example, we have
to deal with a four dimensional torus in a eight dimensional phase
space, which prevents any three dimensional picture. As in the
STIRAP case, the optimal solution presents an oscillatory
structure in the $\theta_2$ variable which is not present in the
adiabatic solution. Following the same intuition as in Sec.
\ref{sec2}, we give up the minimization of the energy and we
choose a cost that allows to minimize the oscillations around
$\theta_2 = \frac{\pi}{2}$:
\begin{equation}\label{eq28}
C = \int_0^T\dot{\theta}_2^2dt = \int_0^T(k \sin \theta_2 \cos
\theta_2 + w_2)^2 dt .
\end{equation}
Plugging this cost into the Hamiltonian and optimizing with
respect to $w_2$, which is the only control in the cost, we get:
$$
w_2 = p_{\theta_2} - k \sin \theta_2  \cos \theta_2,
$$
and the Hamiltonian finally reads:
\begin{equation}\label{eq29}
\begin{array}{rcl}
    H & = & -k r p_r \cos^2 \theta_2 + \frac{1}{2} p_{\theta_2}^2 + w_1 \cot \theta_2 p_{\theta_1}\\
     & & + v_3 \frac{\cot \theta_2 p_{\theta_3}}{\sin \theta_1}.
\end{array}
\end{equation}
It is then straightforward to check from the equations of motion
that the solution $p_{\theta_2}(t) = 0$, for any time $t$,
minimizes the cost $C$. One deduces that $\theta_2$ is constant
along this trajectory and we obtain a relation between $v_3$ and
$w_1$ from $\dot{p}_{\theta_2}=0$:
\begin{equation}
v_3 = (2 k r p_r \cos \theta_2\, \sin^3 \theta_2 - w_1
p_{\theta_1})\frac{\sin \theta_1}{p_{\theta_3}}.
\end{equation}
Hence, by tuning the values of $w_1$, which is assumed here to be
constant, and of the initial momenta, we can reach any final
superposition of the states $|3\rangle$ and $|4\rangle$. For
example, by taking $w_1 = 1$, $r p_r = 100$, $p_{\theta_1}(0) =
16.85$, $p_{\theta_2}(0) = 0, p_{\theta_3}(0) = -1$, we obtain the
same final population in the states $|3\rangle$ and $|4\rangle$ as
shown in Fig. \ref{fig5}. Note the counterintuitive order between
the pump and the Stokes pulses. The corresponding adiabatic
process is associated to the limit $\theta_2\to \pi/2$.
\section{Conclusion and open questions}\label{sec4}
We have reached the main goal of this study, which was to exhibit
a connection between optimal control theory and adiabatic
techniques in quantum systems. A geometric analysis of the
Hamiltonian dynamics constructed from the PMP gives a complete
overview of all the optimal solutions for a given cost. For the
energy minimization problem, we have shown that the adiabatic
pulse can be associated to a peculiar Hamiltonian singularity of
the problem. The adiabatic solution being a singular Hamiltonian
trajectory, it cannot be approached smoothly by any optimal
control field. Therefore, an adapted cost functional has to be
chosen to enforce the optimal solution to follow the structure of
the adiabatic pulse sequence. This study done for three and four
level quantum systems is expected to be generalizable to more
complex quantum dynamics where an adiabatic control scheme can be
used. On the theoretical side, this study is also the first step
in the understanding of the role of Hamiltonian singularities in
control processes. Although much more work need to be done to
advance in this entirely new field, it seems a promising way in
order to design optimal control fields with robustness properties.
For instance, it would be interesting to apply the same idea to
other standard adiabatic processes such as the frequency chirp in
a
two-level quantum system.\\

\noindent \textbf{Acknowledgment}\\
We are grateful to H. Yuan for discussions.

\appendix

\section{Singular reduction theory}\label{app}

In this appendix, we apply the singular reduction theory
\cite{cushman} in order to determine the nature of the singular
tori presented in Sec. \ref{sec2}. We refer the reader to Ref.
\cite{assemat} for a pedagogical introduction to these tools.
Singular reduction theory is a general technique that gives a
global overview of all the Hamiltonian trajectories and of the
corresponding geometrical structures in which they evolve. This
theory is roughly based on the reduction of the dimension of the
phase space by making use of a constant of the motion. However,
caution should be exercised due to the existence of singularities
in the problem. The application of this general procedure to the
example of this work can be summarized as follows.

To simplify the analysis, we work on the dynamics projected on the
sphere, the constant $r \cdot p_r$ playing the role of a
parameter. In cartesian coordinates, this leads to the following
constraints:
\begin{equation}
x_1^2 +x_2^2 + x_3^2 = 1 \quad \textrm{and} \quad x_1 p_{x_1} +
x_2 p_{x_2} + x_3 p_{x_3} = rp_r. \label{eq:AppConstraints}
\end{equation}
We consider the flow of the constant of motion $p_{\phi} = x_1
p_{x_3} - x_3 p_{x_1}$ to reduce the dimension of the problem. We
first introduce the six invariant polynomials under the flow of
$p_{\phi}$, which form a basis for the polynomial functions
$\mathcal{P}$ which Poisson-commute with $p_{\phi}$, i.e.
$\dot{\mathcal{P}}=\left\{ \mathcal{P}, p_{\phi}\right\}=0$. It is
straightforward to check that the following polynomials are
invariant \cite{cushman}:
\begin{equation}
\begin{array}{l}
    \pi_1 = x_2\\
    \pi_2 = p_{x_2}\\
    \pi_3 = x_1 p_{x_3} - x_3 p_{x_1} (= p_{\phi})\\
    \pi_4 = p_{x_1}^2 + p_{x_3}^2\\
    \pi_5 = x_1^2 + x_3^2\\
    \pi_6 = x_1 p_{x_1} + x_3 p_{x_3}\\
\end{array}
\label{eqAppInvpoly}
\end{equation}
These polynomials obey by construction to the reduced phase space
equation:
\begin{equation}
\pi_6^2 + \pi_3^2 = \pi_4 \pi_5. \label{eqAppRedPhaSpa1}
\end{equation}
Since the Hamiltonian $H$ defined in Eq. (\ref{eq15}) commutes
with $p_\phi$, it can be expressed as:
\begin{equation}
H = -k \pi_1 \pi_2 + \frac{1}{2} \left( \pi_1^2 \pi_4 + \pi_2^2 \pi_5 - 2 \pi_1 \pi_2 \pi_6 \right)
\label{eqAppHamRed1}
\end{equation}
Using the two constraints of (\ref{eq:AppConstraints}), Equations
(\ref{eqAppRedPhaSpa1}) and (\ref{eqAppRedPhaSpa2}) become:
\begin{equation}
(rp_r - \pi_1 \pi_2)^2 + \pi_3^2 = \pi_4 (1 - \pi_1^2),
\label{eqAppRedPhaSpa2}
\end{equation}
and
\begin{equation}
H = -(k + rp_r) \pi_1 \pi_2  + \frac{1}{2} \left( \pi_1^2 \pi_4 +
\pi_2^2 \pi_1^2 + \pi_2^2 \right). \label{eqAppHamRed2}
\end{equation}
For a given value of $\pi_3 (= p_{\phi})$, we get two surfaces in
the space $(\pi_1,\pi_2,\pi_4)$. The dynamics takes place at their
intersection. An example of such intersection is plotted in Fig.
\ref{fig3}. This eight-like shape is the mark of a specific type
of singular tori, the bitorus which is represented in Fig.
\ref{fig1}.

\end{document}